\documentclass[preprint,proceedings]{rmaa}

\suppressfulladdresses 

\SetYear{2007}
\SetConfTitle{Massive Stars: Fundamental Parameters and Circumstellar Interactions}

\title{Jet interactions in massive X-ray binaries} 

\author{
  Gustavo E. Romero \altaffilmark{1,2} 
  }

\altaffiltext{1}{Facultad de Ciencias Astron\'omicas y Geof\'{\i}sicas, Universidad Nacional de La Plata, Paseo del Bosque, 1900 La Plata, Argentina (romero@fcaglp.unlp.edu.ar).}

\altaffiltext{2}{Instituto Argentino de Radioastronom\'{\i}a, Casilla de Correos No. 5, (1894) Villa Elisa, Buenos Aires, Argentina (romero@irma.iar.unlp.edu.ar).}

\shortauthor{Gustavo E. Romero}
\shorttitle{Jet interactions in HMXRBs}

\listofauthors{Gustavo E. Romero}
\indexauthor{Romero, G.E.}

\abstract{Massive X-ray binaries are formed by a compact object that accretes matter from the stellar wind of an early-type donor star. In some of these systems, called microquasars, relativistic jets are launched from the surroundings of the compact object. Such jets interact with the photon field of the companion star, the stellar wind, and, at large distances, with the interstellar medium. In this paper I will review the main results of such interactions with particular emphasis on the production of high-energy photons and neutrinos. The case of some specific systems, like LS~I~+61~303, will be discussed in some detail. Prospects for future observations at different wavelengths of this type of objects will be presented.}

\resumen{Los sistemas binarios masivos de rayos X est\'an formados por un objeto compacto que acreta materia del viento estelar de una estrella donante de tipo temprano. En algunos de estos sistemas, llamados microcu\'asares, chorros de part\'\i{}culas relativistas son eyectados desde las cercan\'\i{}as del objeto compacto. Estos chorros interact\'uan con el campo de fotones de la estrella compa\~nera, con el viento estelar, y, a grandes distancias, con el medio interestelar. En este trabajo se resumir\'an los principales resultados de tales interacciones con especial \'enfasis en la producci\'on de fotones de alta energ\'\i{}a y neutrinos. El caso de alg\'un sistema particular, como ser LS~I~+61~303, ser\'a discutido con alg\'un detalle. Adem\'as, se presentar\'an las perspectivas futuras para observaciones a diferentes longitudes de onda para este tipo de objetos.}

\addkeyword{Gamma rays: theory}
\addkeyword{Gamma rays: observations}
\addkeyword{Jets and outflows}
\addkeyword{Stars: binaries}
\addkeyword{Stars: microquasars}

\begin{document}
\maketitle

\section{Introduction}
\label{sec:intro}
Massive stars use to form binary systems. In such systems one of the stars evolves faster than the other. At the end of the lifetime of this star a supernova explosion will occur, and either a neutron star or a black hole will be left behind. If the system is not disrupted by the explosion, the compact object will start to accrete matter from the stellar wind of its early-type companion. Since the matter has angular momentum, it will form an accretion disk around the compact star. The matter will be heated in the disk, losing angular momentum and falling into the potential well. The hot disk will cool through the emission of X-rays. We say then that a massive X-ray binary (HMXRB) is born. There are around 120 HMXRBs detected in the Galaxy so far (Liu et al. 2006). Some of these systems present non-thermal radio emission. This emission is thought to be synchrotron radiation produced by relativistic electrons in a jet that is somehow ejected from the surroundings of the compact object. When the jet is resolved at radio wavelengths through interferometric techniques or at X-rays, the HMXRB is called a high-mass {\sl microquasar} (Mirabel et al. 1992). 

The word `microquasar' (MQ) was coined to emphasize the similarities between galactic jet sources and extragalactic quasars (Mirabel \& Rodr\'\i guez 1998). These similarities, although important, should not make us to overlook the also important differences between both types of objects. The main difference is, of course, the presence of a donor star in the case of MQs. In high-mass MQs, this star provides a strong photon field, a matter field in the form of a stellar wind, and a gravitational field that can act upon the accretion disk producing a torque and inducing its precession. The photon and matter field constitute targets for the relativistic particles in the jet. The interaction of the jets with these fields can produce a variety of phenomena that are absent in the case of extragalactic jets. The aim of the present article is to review these phenomena.

\section{What is a microquasar?}
\label{sec:micro}

\begin{figure}[!t]
  \includegraphics[angle=-90,width=1\columnwidth]{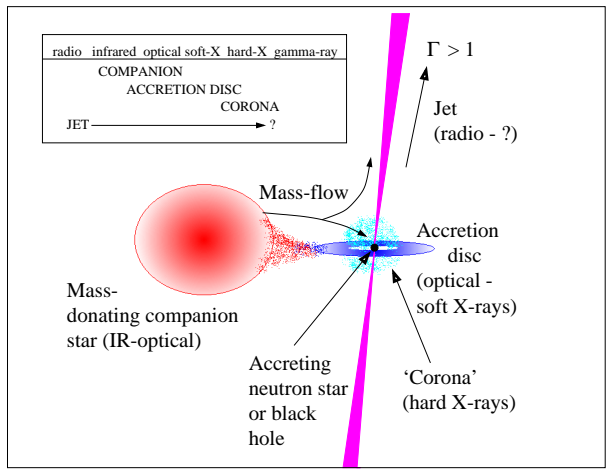}
  \caption{Sketch showing the different components of a microquasar and the energy bands at which they emit. Not to scale. From Fender \& Maccarone (2004).}  
  \label{fig:sketch}
\end{figure}

A microquasar is an accreting X-ray binary system with non-thermal jets. The basic ingredients of a MQ are shown in Figure \ref{fig:sketch}. They are the compact object, the donor star, the accretion disk, the jets, which usually are relativistic or mildly relativistic, and a region of hot plasma called the `corona' that surrounds the compact object. If the star is an early-type, hot star, the accretion can proceed through capture of the wind material. In the case of low-mass stars and in some close systems, the accretion occurs through the overflow of the Roche lobe. In what follows we will focus only on high-mass MQs.   

The donor star can produce radiation from the IR up to UV energies. The accretion disk produces soft X-rays, whereas the corona is responsible for hard X-rays that are likely generated by Comptonization of disk photons. The emission of the jets goes from radio wavelengths to, in some cases like LS 5039, gamma-rays. MQs, like blazars, can emit along the entire electromagnetic spectrum. Their spectral energy distribution (SED) is complex, being the result of a number of different radiative processes occurring on different size-scales in the MQs. 

MQs present different spectral states at X-rays. The two basic state are the `soft' state and the `hard' state. In the former the SED is dominated by a greybody peak around $E\sim1$ keV, probably due to the contribution of the accretion disk, which extends in this state all the way down to the last stable orbit around the compact object. In the hard state the peak in the X-rays is shifted toward lower energies and a strong and hard power-law component is present up to energies $\sim 150$ keV, in some cases even beyond. This emission is usually interpreted as soft X-ray Comptonization in the corona (e.g. Ichimaru 1977), although some authors have suggested that it could be produced in the jet through external inverse Compton (IC) interactions (Georganopoulos et al. 2002) or through synchrotron mechanism (Markoff et al. 2001, 2003).

The sources spend most of the time in the hard state. It is in this state when a steady, self-absorbed radio jet is usually observed. The transition form one state to the other is commonly accompanied by the ejection of superluminal components, that can be detected as moving radio blobs (Mirabel \& Rodr\'\i guez 1994, Fender et al. 2004).           

\section{What are jets made of?}
\label{jets}

One of the most important open issues concerning MQs is the nature of the matter content of the jets. We know for sure that relativistic leptons with a power-law distribution are present in the jets since we can detect and measure their synchrotron radiation. The relativistic outflow can be made of relativistic electron-positron pairs, or alternatively it could be a relativistic proton-electron plasma. Another possibility is a plasma formed by a cold electron-proton fluid, where the particles would have a thermal distribution, plus a relativistic content, locally accelerated by shocks (Bosch-Ramon, Romero \& Paredes 2006). In this kind of jets, the bulk of the momentum is carried out by the cold plasma, which additionally confines the relativistic component. 

In any case, the large perturbations observed in the interstellar medium (ISM) around some MQs like Cygnus X-1 (Gallo et al. 2005) and SS 433 (Dubner et al. 1998), strongly suggest that the jets are barionic loaded. The direct detection of iron lines in the case of the jets of SS 433 (Kotani et al. 1994, 1996; Migliari et al. 2002) clearly confirms that they contain hadrons, at least in this particular object. Since there seems to be a clear correlation between the accretion and ejection of matter in MQs (Mirabel et al. 1998), it is natural to assume that the content of the jets does not basically differ in nature from that of the accreting matter. All these considerations make quite likely the presence of relativistic protons in the jets of MQs. Hence, their radiative signatures can not be neglected in a serious analysis of the radiative processes in these sources.  

\section{Jet interactions}
\label{sec:int}

What does happen when a relativistic jet pass through the medium that surrounds a hot, massive star?. The radiation field of the star penetrates freely into the jet and the dominant UV photons will interact with relativistic particles in the outflow. The interaction of the stellar wind with the jet will form a boundary layer where shocks will likely be formed, but some level of fluid mixing is expected to occur. The interaction between relativistic particles from the jet and thermal particles of the wind will take place, producing high-energy emission. We can separate the microscopic jet-stellar environment interactions in two groups, according to whether they are of leptonic or hadronic nature. Of course, both types of reactions will occur in a specific system, but according to the given conditions, one type or the other might dominate the high-energy output of the MQ. Let us briefly discuss both cases.
      
\subsection{Leptonic interactions}

Relativistic electrons and positrons in the jet will IC scatter soft photons up to high energies. The origin of these photons can be diverse: stellar UV photons, X-ray photons from the accretion disk and the hot corona around the compact object, or non-thermal photons produced in the jet by synchrotron mechanism. At high energies, the interaction enters in the Klein-Nishina regime, where the cross section decreases dramatically. Opacity effects to gamma-ray propagation due to the presence of the local photon fields can result in the generation of IC cascades within the binary system (Bednarek 2006a, Orellana et al. 2007). Relativistic leptons can interact with cold protons and nuclei from the stellar wind producing high-energy emission through relativistic Bremsstrahlung. A number of papers have been devoted to leptonic interactions in MQs in recent years, for instance, Atoyan \& Aharonian (1999), Markoff et al. (2001, 2003), Georganopolous et al. (2002), Kaufman Bernad\'o et al. (2002), Romero et al. (2002), Bosch-Ramon \& Paredes (2004), Bosch-Ramon et al. (2005a, 2006), Paredes et al. (2006), Dermer \& B\"ottcher (2006), Gupta et al. (2006), Bednarek (2006b), etc. The reader is referred to these papers and references therein for detailed discussions.

\begin{figure}[!t]
  \includegraphics[angle=0,width=1\columnwidth]{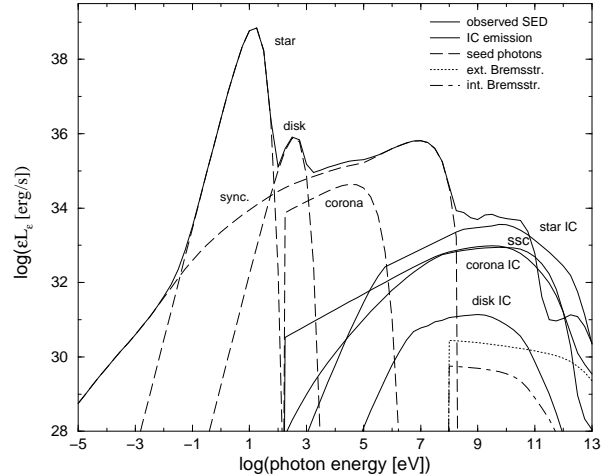}
  \caption{Spectral energy distribution of a high-mass MQ. The different contributions to the total SED are shown.   From Bosch-Ramon, Romero \& Paredes (2006).}
  \label{fig:SED1}
\end{figure}

\begin{figure}[!t]
  \includegraphics[angle=0,width=1\columnwidth]{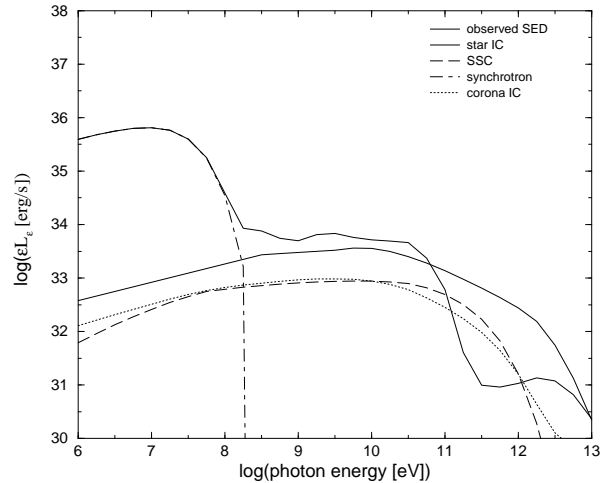}
  \caption{High-energy emission from high-mass MQ. Courtesy of V. Bosch-Ramon. }
  \label{fig:SED2}
\end{figure}

In Figure \ref{fig:SED1} we show the broadband SED expected from leptonic interactions in a high-mass MQ. The different contributions are indicated. It can be seen that the synchrotron emission can extend up to MeV energies and that in the GeV-TeV range the dominant process is IC upscattering of stellar photons. Figure \ref{fig:SED2} shows a detail of the SED at high energies. Notice that absorption by photon-photon annihilation has been taken into account in the final curve, yielding a soft spectrum around 100 GeV (Bosch-Ramon et al. 2006).

\subsection{Hadronic interactions}

The main reaction for proton cooling in a high-mass MQs is $pp$ interaction, through the channels $pp\rightarrow p+p+\pi^{0}$ and $pp\rightarrow p+p+\xi(\pi^{+}+\pi^{-})$, where $\xi$ is the $\pi^{\pm}$ multiplicity. The neutral pions decay yielding gamma rays, $\pi^{0}\rightarrow\gamma+\gamma$, whereas the charged pions produce neutrinos and $e^{\pm}$ pairs: $\pi^{\pm}\rightarrow\mu^{\pm}\nu_{\mu}\bar{\nu}_{\mu}\rightarrow e^{\pm}\nu_{e}\bar{\nu}_{e}$. The gamma-ray spectrum will mimic at high-energies the spectrum of the parent relativistic proton population. In general, since proton losses are not as severe as electron losses in the inner region of the source, we could expect a higher energy cutoff in hadronic-dominated sources.

Models for hadronic MQs have been developed by Romero et al. (2003), Romero et al. (2005), Romero \& Orellana (2005) and Orellana \& Romero (2007). Neutrino production in this kind of models is discussed by Romero \& Orellana (2005), Aharonian et al. (2006), Benarek (2005) and Christiansen et al (2006). For photo-pion production of neutrinos, under rather extreme conditions, see Levinson \& Waxman (2001). 

 \begin{figure}[!t]
  \includegraphics[angle=-90,width=1\columnwidth]{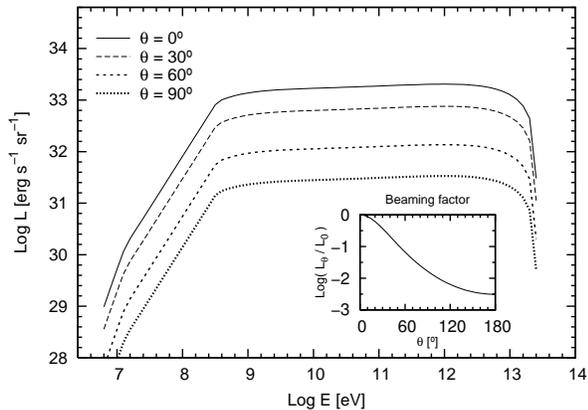}
  \caption{Spectral energy distributions for the hadronic emission of a high-mass MQ with a smooth stellar wind. Different curves correspond to different viewing angles. From Orellana et al. (2007).}
  \label{fig:hadron}
\end{figure}

Figure \ref{fig:hadron} shows different SED obtained from $pp$ interactions for a high-mass MQ with a smooth spherical wind (Orellana et al. 2007). The various curves correspond to different viewing angles. The total jet power in relativistic protons is $L_p^{\rm rel}=6\times 10^{36}$ erg s$^{-1}$. The jet is assumed to be perpendicular to the orbital plane, but this constraint can be relaxed to allow, for instance, for a precessing jet. Actually, in some systems, the jet could point even in the direction of the star (Butt et al. 2003, Romero \& Orellana 2005). In such a case, jet-induced nucleosynthesis can occur in the stellar atmosphere. The power of the stellar wind might be, for some stars, strong enough as to stop the jet creating a standing shock between the compact object and the star. Protons and electrons might be re-accelerated there up to very high energies, producing a detectable TeV source. 

All existing models for hadronic MQs assume a smooth wind from the star\footnote{See, nonetheless, the paper by Aharonian \& Atoyan (1996) that, although not framed in the context of MQs, discusses the interaction of a proton beam with a cloudy medium around a star.}. However, it would be quite possible that the wind have some structure, for instance in the form of clumps, a fact that would lead to gamma-ray variability on short time scales. If such a variability could be detected by future Cherenkov telescope arrays, it might be used to infer the structure of the wind. The jet would act as a kind of lantern illuminating the wind in gamma-rays to the observer.  

Hadronic jets can propagate through the ISM producing hot spots similar to those observed in the case of extragalactic sources (Bosch-Ramon et al. 2005b). Particles re-accelerated at the termination point of the jets, can diffuse in the ambient medium, interacting with diffuse material and producing extended high-energy sources.

\section{The controversial case of LS~I~+61~303}

\begin{figure}[!t]
  \includegraphics[angle=0,width=1\columnwidth]{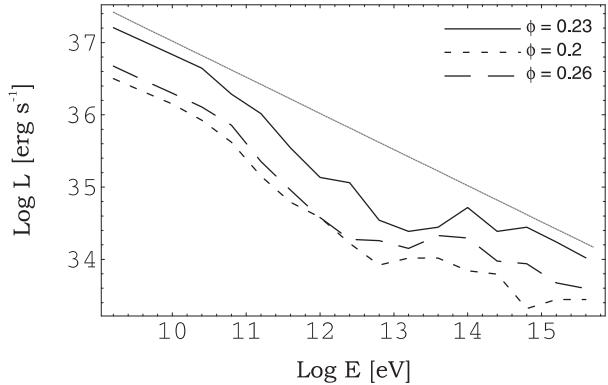}
  \caption{Electromagnetic cascades at different phases developed close to the periastron passage of the X-ray binary LS~I~+61~303 (from Orellana \& Romero 2007).}
  \label{fig:cascades}
\end{figure}

LS~I~+61~303 is a puzzling Be/X-ray binary, which displays gamma-ray variability at high energies. The nature of the compact object and the origin of the high-energy emission is unclear. The detection of jet-like radio features by Massi et al. (2001, 2004) led to the classification of this source as a MQ. This has been recently challenged by Dhawan et al. (2006), who observed the source with the VLBA at different orbital phases concluding that the direction the jet-like feature during the periastron passage (opposed to the primary star) supports the scenario of a colliding wind model where the compact object is an energetic pulsar (wind power $\sim 10^{36}$ erg/s). The system has been detected by the MAGIC telescope at $E>200$ GeV. The variability is modulated with the orbital period. Contrary to the expectations the maximum of the gamma-ray emission occurred well after the periastron passage. The source was not clearly detected during the periastron (Albert et al. 2006). The cause of this could be gamma-ray absorption in the combined photon field of the Be star and its decretion disk (e.g. Orellana \& Romero 2007). Figure \ref{fig:cascades} shows the electromagnetic cascades that might develop close to the periastron passage (which occurs at phase 0.23). According to these simulations the source should be detectable during the periastron, but with longer exposures, and the spectrum will be softer than what was observed at phases 0.6-0.7. 

The pulsar/Be binary interpretation goes not free of severe problems. The flux at MeV-GeV energies observed by EGRET (Kniffen et al. 1997) accounts for a luminosity of $\sim 10^{36}$ erg/s, which would imply an impossible conversion efficiency from wind power to gamma-rays of $\sim 1$. In addition, since the pulsar wind would be orders of magnitude stronger than the slow Be equatorial wind, the observed `cometary tail' radio feature, if interpreted as synchrotron radiation from electrons accelerated at the colliding wind region, should point out {\sl toward} the primary star, and not opposite to it. 

It is clear that LS~I~+61~303 is a interesting and peculiar system that deserves more intensive studies in the near future.

\section{Conclusions}

MQs are outstanding natural laboratories to study a variety of physical phenomena such as particle acceleration, accretion physics, and particle interactions. Observations of gamma-ray emission of high-mass MQs can be used to probe the structure of stellar winds and the nature of the matter content of relativistic jets. 

How many MQs are there in the Galaxy?. It is difficult to answer this questions, but it seems possible that a significant number of the yet-unidentified variable gamma-ray sources located on the galactic plane (Romero et al. 1999) could be associated with high-mass MQs (Romero et al. 2004, Bosch-Ramon et al. 2005a). In the next few years, new Cherenkov telescope arrays like HESS II, MAGIC II, and VERITAS, along with the satellite observatories AGILE and GLAST, will continue detecting these extraordinary objects at high energies and helping to penetrate into their mysteries.    

\section*{Acknowledgments}
This work has been supported by the Agencies CONICET (PIP 5375) and ANPCyT (PICT 03-13291 BID 1728/OC-AR). I thank the organizers for a wonderful meeting and a warm hospitality.


\begin{thebibliography}

\bibitem{}Albert, J. et al. (MAGIC coll.) 2006, Science, 312, 1771
\bibitem{}Aharonian, F. A., \& Atoyan, A. M. 1996, Space Sci. Rev., 75, 357
\bibitem{}Aharonian, F. A., Anchordoqui, L. A., Khangulyan, D., \& Montaruli, T. 2006, Journal of physics: conference series, 39, 408
\bibitem{} Atoyan, A.~M., \& Aharonian, F.~A. 1999, MNRAS, 302, 253
\bibitem{}Bednarek, W. 2005, ApJ, 631, 466
\bibitem{}Bednarek, W. 2006a, MNRAS, 368, 579
\bibitem{}Bednarek, W. 2006b, MNRAS, 371, 1737
\bibitem{}Bosch-Ramon, V.~\& Paredes, J.~M.\ 2004, A\&A, 417, 1075 
\bibitem{}Bosch-Ramon, V., Romero, G.~E., \& Paredes, J.~M. 2005a, A\&A, 429, 267
\bibitem{}Bosch-Ramon, V., Aharonian, F. A., \& Paredes, J. M. 2005b, A\&A, 432, 609
\bibitem{}Bosch-Ramon, V., Romero, G.~E., \& Paredes, J.~M. 2006, A\&A, 447, 263
\bibitem{}Butt, Y.M., Maccarone, T.J., \& Prantzos, N. 2003, ApJ, 587, 748
\bibitem{}Christiansen, H. R., Orellana, M., \& Romero, G. E. 2006, PhRvD, 73, 063012 
\bibitem{} Dhawan, V., Mioduszewski, A., \& Rupen, M., 2006, in {Proc. of the VI Microquasar Workshop}, Como-2006
\bibitem{}Dermer, C., \& B\"ottcher, M. 2006, ApJ, 643, 1081 
\bibitem{}Dubner, G.~M., Holdaway, M., Goss, W.~M., \& Mirabel, I.~F. 1998, AJ, 116, 1842
\bibitem{}Fender R., \& Maccarone T. 2004, in: Cosmic Gamma-Ray Sources, ed. K.S. Cheng \& G.E. Romero, Kluwer
Academic Publishers, Dordrecht, p.205
\bibitem{}Fender, R. P., Belloni, T. M., \& Gallo, E. 2004, MNRAS, 355, 1105
\bibitem{}Gallo, E., Fender, R., Kaiser, C. 2005, Nature, 436, 819
\bibitem{}Georganopoulos, M., Aharonian, F. A., \& Kirk, J. G.
2002, A\&A, 388, L25
\bibitem{}Gupta, S., B\"ottcher, M., \& Dermer, C. D. 2006, ApJ, 644, 409
\bibitem{}Ichimaru, S. 1977, ApJ, 214, 840
\bibitem{}Kaufman Bernad\'o, M. M., Romero, G. E., \& Mirabel, I. F. 2002, A\&A, 385, L10
\bibitem{}Kniffen, D.A., et al., 1997, ApJ, 486, 126 
\bibitem{}Kotani, T., Kawai, N., Aoki, T., et al. 1994, PASJ, 46, L147
\bibitem{}Kotani, T., Kawai, N., Matsuoka, M., \& Brinkmann, W. 1996, PASJ, 48, 619
\bibitem{}Levinson, A., \& Waxman, E. 2001, PhRvL, 87, 171101
\bibitem{}Liu, Q.Z., van Paradijs, J., \& van den Heuvel, E. P. J 2006, A\&A, 455, 1165 
\bibitem{}Markoff, S., Falcke, H., \& Fender, R. P. 2001, A\&A, 372, L25
\bibitem{}Markoff, S., Nowak, M., Corbel, S., et~al. 2003, A\&A, 397, 645 
\bibitem{}Massi, M., et al. 2001, A\&A, 376, 217  
\bibitem{}Massi, M., et al. 2004, A\&A, 414, L1
\bibitem{}Migliari, S., Fender, R. \& M\'endez, M. 2002, Science, 297, 1673
\bibitem{}Mirabel, I. F., Rodriguez, L. F., Cordier, B., Paul, J., \& Lebrun, F. 1992, 
Nature, 358, 215
\bibitem{}Mirabel, I. F., \& Rodr\'{\i}guez, L. F. 1994, Nature, 371, 46
\bibitem{}Mirabel, I. F., \& Rodr\'{\i}guez, L. F. 1998, Nature, 392, 673
\bibitem{}Mirabel, I. F., Dhawan, V., \& Chaty, S. et~al. 1998, A\&A, 330, L9
\bibitem{}Orellana, M., \& Romero, G. E. 2007, Ap\&SS, in press
\bibitem{}Orellana, M., Bordas, P., Bosch-Ramon, V., et al. 2007, A\&A, submitted 
\bibitem{}Paredes, J. M., Bosch-Ramon, V., \& Romero, G. E. 2006, A\&A, 451, 259
\bibitem{}Romero, G.E., Benaglia, P., Torres, D.F. 1999, A\&A, 348, 868
\bibitem{}Romero, G.E., Kaufman Bernad\'o, M.M., \& Mirabel, I.F. 2002,
A\&A, 393, L61
\bibitem{}Romero, G.~E., Torres, D.~F., Kaufman  Bernad\'o, M.~M., \& Mirabel, I.~F. 2003, A\&A, 410, L1  
\bibitem{}Romero, G.~E., Grenier, I.~A., Kaufman Bernad\'o, M.M., Mirabel, I.F., \& Torres, D.~F. 
2004, ESA-SP, 552, 703
\bibitem{}Romero, G.E., \& Orellana, M. 2005, A\&A, 439, 237
\bibitem{}Romero, G.E., Christiansen, H.R., \& Orellana, M. 2005 ApJ, 632, 1093
 
\end{thebibliography}
\end{document}